\documentclass[iop]{emulateapj}
\usepackage{url}

\usepackage{soul}
\usepackage{rotating}
\usepackage{threeparttable}
\shorttitle{Evolution of Metals in MgII absorbers}
\shortauthors{Lan \& Fukugita}

\begin{document}
\title{M\MakeLowercase{g}\,II absorbers: metallicity evolution and cloud morphology}

\author{Ting-Wen Lan$^{1}$ and Masataka Fukugita$^{1,2}$}

\affil{$^{1}${Kavli Institute for the Physics and Mathematics of the Universe,
    University of Tokyo, Kashiwa, Chiba 277-8583, Japan}}
\affil{$^{2}${Institute for Advanced Study, Princeton, NJ 08540, U.S.A.}}

\begin{abstract}
  Metal abundance and its evolution are studied for Mg II quasar
  absorption line systems from their weak, unsaturated spectral lines
  using stacked spectra from the archived data of Sloan Digital Sky
  Survey. They show an abundance pattern that resembles that of the 
  Galactic halo or Small Magellanic Cloud, with metallicity [Z/H]
  showing an evolution from redshift $z=2$ to 0.5: metallicity becomes
  approximately solar or even larger at $z\approx0$.  We show that the
  evolution of the metal abundance traces the cumulative amount of the
  hydrogen fuel consumed in star formation in galaxies.  With the aid
  of a spectroscopic simulation code, we infer the median gas density of
  the cloud to be roughly 0.3~${\rm cm^{-3}}$, with which the
  elemental abundance in various ionization stages, in particular C I, is
  consistently explained.  This gas density implies that the size of
  the Mg II clouds is of the order of 0.03~kpc, which suggests that
  individual Mg II clouds around a galaxy are of a baryonic mass typically $10^3
  \rm \, M_{\odot}$. This means that Mg II clouds are numerous and `foamy',
  rather than a large entity that covers a sizable fraction of
  galaxies with a single cloud.
\end{abstract}

\keywords{quasars: absorption lines, galaxies: haloes}

\section{Introduction}

Mg II quasar absorption clouds ubiquitously reside in the vicinity of
galaxies, typically within their virial radii, in circumgalactic space
\citep{Bergeron1991, Steidel1994}.  The absorption features at intervening redshifts are detected from 35-40\% of quasar spectra \citep[e.g.,][for modern SDSS data]{Zhu2013} and MgII clouds cover as large as 50\% of the sky around galaxies   typically at redshift $\sim0.5$ \citep[e.g.][]{Chen2010,Menard2012,Nielsen2013,Lan2014}.  We expect  the clouds affected by neighbouring galaxies, yet their nature and formation are not known
well, for our knowledge of the cloud is limited to line of sight
observations. The feature we now know is that MgII clouds are
significantly contaminated with metals, habouring dust
\citep[e.g.][]{York2006,Menard2012}, whereas star formation activity
therein is not known, nor expected given the low column density of
the clouds. The morphology of the clouds is also yet to be known: whether they consist of
several large clouds that cover a significant fraction of galaxies, or
an assembly of small clouds.

Elemental analyses for these clouds are often hampered by the fact
that important metal lines are saturated, whereas weak lines suffer
from poor signal to noise ratios to carry out a detailed analysis.  In
the present work we explore metal lines, using weak, unsaturated lines
by stacking many spectra to give sufficient signal to noise ratios for
those absorption lines.  This enables us to derive elemental abundance
as average quantities, as a function of the strength of spectral lines
and redshift specified, and then the pattern of the elemental
abundance, and/or its evolution.

One can also infer the physical state of the gas with the aid of a spectroscopic synthesis code, CLOUDY
\citep{Ferland2013}, which reveals the physical state of the gas cloud
for a given number density of atoms, metallicity and ionising
fields. This, after verifying the validity of the code for the
problems that concern us, would lead us to infer physics of the
clouds. This in turn hints us to infer their morphology around
galaxies.

Our data analysis is written in Section 2 to measure the equivalent
width, with some results in our intermediate steps
given in Appendix. We
discuss in Section 3 the column density of elements that show
absorption features in MgII clouds.  The CLOUDY code was employed to
infer the physical state of the gas in the clouds. In this work all
equivalent widths referred to are those in the rest frame. Wherever we
refer to the average value, we take median quantities to avoid largely
deviated data unless otherwise explicitly stated.  When we refer to
the solar composition, it is that by \citet{Asplund2009}.  Section
4 is given for the summary of our analysis. We use $H_0=70$ km\,s$^{-1}$Mpc$^{-1}$,
and $\Omega_{\rm M}=0.3$ in a flat Universe. $W_{\lambda 2796}$ refers to the rest equivalent width of MgII $\lambda 2796$ line.

\section{Data analysis}

\subsection{Composite spectra for metal absorption lines}

We use the metal absorber
catalogue\footnote{\url{http://www.guangtunbenzhu.com/jhu-sdss-metal-absorber-catalog}}
and the corresponding spectra compiled by \citet{Zhu2013} from quasar
spectra of the Sloan Digital Sky Survey I-III \citep{SDSS2000}.  The
sample contains 77,647 MgII absorbers from redshift 0.4 to 2.5,
detected in 142,012 quasar spectra from the DR7 \citep{DR7quasar} and
DR12 \citep{DR12quasar} quasar catalogues. In the present study, we
take 70,713 systems with $W_{\lambda 2796}>0.4 \, \rm \AA$.
The completeness of MgII absorbers drops gradually from $W_{\lambda 2796}=0.8 \, \rm \AA$ and it is about $30\%$ at
$W_{\lambda 2796}=0.4\, \rm \AA$. The completeness is not
an important issue in major parts of our analysis.

\begin{figure}
\centering
\includegraphics[scale=0.43]{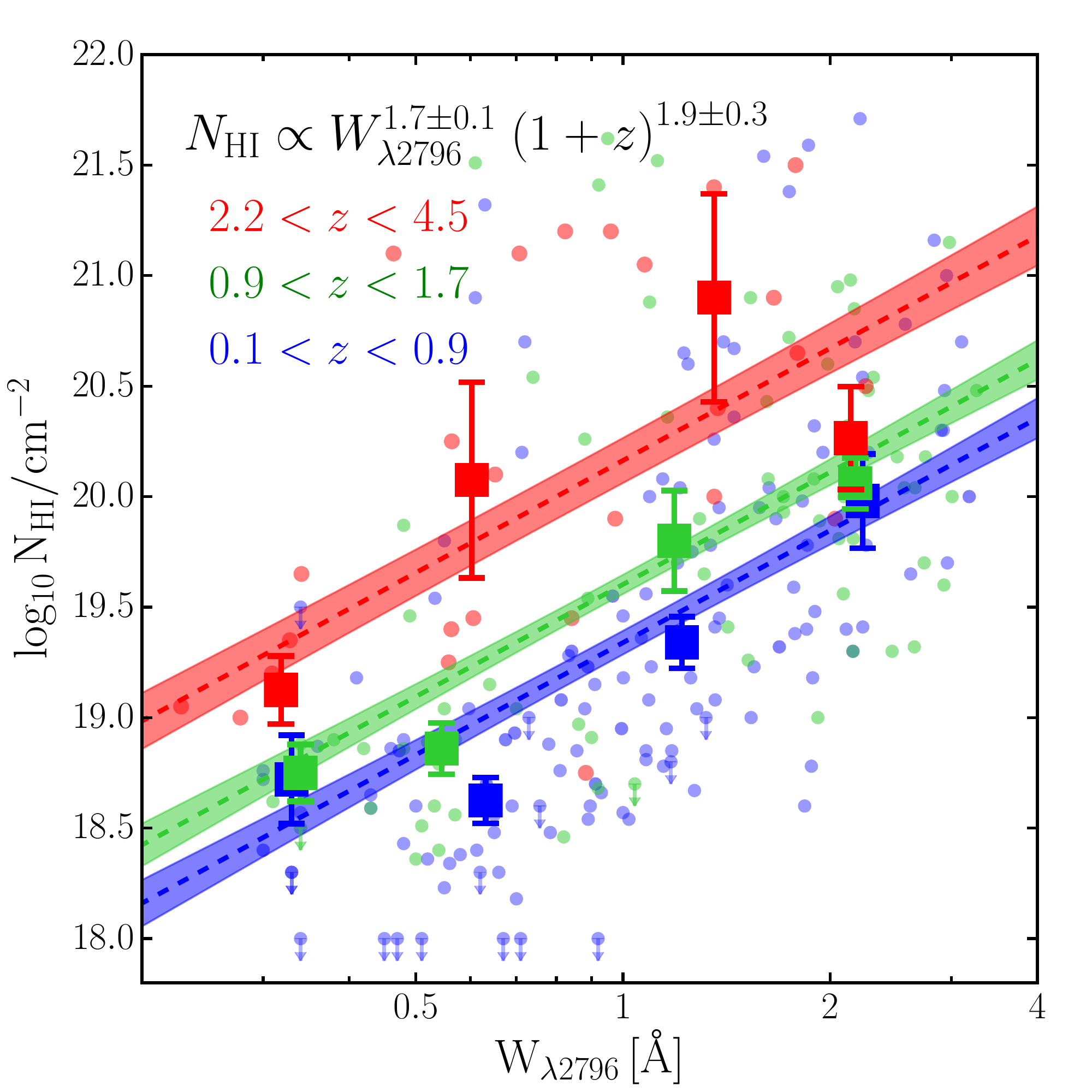}
\caption{$N_{\rm HI}$ versus MgII equivalent widths of the clouds. 
Symbols with three colours represent samples in different redshifts,
as shown in the legend with the redshift ranges presented. Square data points are median values of the three samples with bootstrapping errors. Three lines are fit
  to the data with equation (2) using common $\alpha$ and $\beta$,
  as explained in the text.}
\label{}
\end{figure}


We make median composite spectra of MgII absorbers divided into bins
of their absorption strengths, as characterized with MgII$\lambda2796$,
and redshifts. We focus on rest-frame wavelengths longer than 1250
$\rm \AA$. The median estimator is used to avoid too strong effects of outliers in
spectra that occasionally occur.  We emphasize that it is essential to
our work to measure weak unsaturated absorption lines accurately.

The metal absorption lines we measure are listed in Table 1 below. The
typical signal-to-noise ratio of the composite spectra per spectral
resolution element (70 km/s) is about 500 that allows us to measure
absorption features to the level of $0.01 \rm \AA$ in equivalent
widths.

We measure the rest-frame equivalent widths of isolated metal lines
with a single Gaussian profile fitting. For wavelength regions with
multiple lines crowded, we adopt multiple Gaussian profiles to fit all
relevant absorption lines in both vicinity of lines and continuum.  A special treatment is used to obtain the rest equivalent widths of ZnII to deblend lines of CrII and MgI \citep[e.g.,][]{York2006}.  The errors
of the rest equivalent widths are estimated by bootstrapping the
sample for 200 times. In Appendix, we show an example of the composite spectra, the Gaussian profile fitting and the measured rest equivalent widths of metal lines as a function of redshift and MgII strengths.

For weak absorption lines we can estimate the column density with the
linear relation of the curve of growth \citep[e.g.,][]{Draine2011},
\begin{equation}
N_{\rm ion} \ [{\rm cm^{-2}}] = 1.13\times 10^{20}\times\frac{W_{\rm
    ion}}{f \lambda^{2}},
\end{equation}
where the oscillator strength $f$ is from \citet{Morton2003} and
$\lambda$ (\AA) the rest-frame wavelength of the absorption line. For
weak unsaturated lines we give in Table 1 (Column 3) the estimated
column density.  By comparing measurements with the theoretical curve
of growth, we confirm that those absorption lines are in the linear
regime.  The strong, saturated absorption lines are marked in Column
(3) with an asterisk.

\begin{table*}[ht] 

\caption{Best fit parameters for Equation 3, $W_{\lambda}=C(W_{\lambda 2796})^{\alpha}(1+z)^{\beta}$}
\centering
\begin{threeparttable}
  \begin{tabular}{| c | c | c | c | c | c |}
\hline\hline
\hspace{0.4cm}
Transition & wavelength  $[\rm \AA]$ &  $N$ column density$^{a}$ [$10^{12}\, \rm cm^{-2}$] & $\alpha$ & $\beta$ & $C$ [$\rm \AA$]\\
\hline\hline
HI & $1215.67$ &  $10^{6.96\pm0.10}$ & $1.69\pm0.13$ & $1.88\pm0.29$  & \\

&&&&&\\
SiII & $1260.42$ & *Saturated$^{b}$& $1.07\pm0.01$  & $-0.22\pm0.13$ & $0.573\pm0.086$ \\
OI & $1302.17$ & * & $1.48\pm0.02$  & $0.28\pm0.17$ & $0.145\pm0.028$ \\
SiII & $1304.37$ & * & $1.35\pm0.03$  & $-0.62\pm0.19$ & $0.325\pm0.069$ \\
NiII & $1317.22$ &  $0.4\pm1.0$ & $1.53\pm0.29$ & $3.25\pm2.16$ & $0.000\pm0.001$ \\
CII & $1334.53$ & * & $1.05\pm0.01$  & $-0.27\pm0.07$ & $0.635\pm0.051$ \\
NiII & $1370.13$ &  $42.8\pm55.5$ & $1.45\pm0.18$ & $-1.41\pm1.14$ & $0.055\pm0.071$ \\
SiIV & $1393.76$ & * & $0.66\pm0.01$  & $-1.01\pm0.10$ & $1.124\pm0.122$ \\
SiIV & $1402.77$ & * & $0.65\pm0.02$  & $-1.05\pm0.13$ & $0.793\pm0.111$ \\
NiII & $1454.84$ &  $23.3\pm49.4$ & $1.60\pm0.36$ & $-0.99\pm2.00$ & $0.014\pm0.030$ \\
SiII & $1526.71$ & * & $1.35\pm0.01$  & $-0.25\pm0.05$ & $0.313\pm0.016$ \\
CIV & $1548.20$ & * & $0.58\pm0.01$  & $-0.87\pm0.05$ & $1.524\pm0.081$ \\
CIV & $1550.78$ & * & $0.60\pm0.01$  & $-1.30\pm0.06$ & $1.728\pm0.104$ \\
CI & $1560.31$ &  $56.6\pm65.6$ & $1.99\pm0.36$ & $-3.09\pm1.10$ & $0.094\pm0.109$ \\
FeII & $1608.45$ & * & $1.70\pm0.02$  & $-0.76\pm0.09$ & $0.165\pm0.016$ \\
CI & $1656.93$ &  $124.3\pm55.2$ & $1.97\pm0.13$ & $-3.86\pm0.45$ & $0.449\pm0.200$ \\
AlII & $1670.79$ & * & $1.33\pm0.01$  & $-0.33\pm0.03$ & $0.375\pm0.012$ \\
NiII & $1709.60$ &  $9.5\pm4.4$ & $1.48\pm0.11$ & $0.11\pm0.45$ & $0.008\pm0.004$ \\
NiII & $1741.55$ &  $23.6\pm6.7$ & $1.52\pm0.07$ & $-0.70\pm0.28$ & $0.027\pm0.008$ \\
NiII & $1751.92$ &  $29.8\pm11.1$ & $1.45\pm0.10$ & $-0.78\pm0.36$ & $0.022\pm0.008$ \\
SiII & $1808.01$ &  $562.3\pm89.1$ & $1.67\pm0.04$ & $-0.50\pm0.15$ & $0.034\pm0.005$ \\
AlIII & $1854.72$ & * & $1.19\pm0.01$  & $-0.85\pm0.05$ & $0.255\pm0.012$ \\
AlIII & $1862.79$ &  $16.8\pm1.3$ & $1.24\pm0.02$ & $-0.86\pm0.08$ & $0.143\pm0.011$ \\
ZnII & $2026.14$ &  $1.4\pm0.3$ & $1.96\pm0.08$ & $-1.10\pm0.20$ & $0.025\pm0.005$ \\
CrII & $2056.26$ &  $5.8\pm1.2$ & $1.47\pm0.08$ & $-0.69\pm0.23$ & $0.022\pm0.005$ \\
ZnII & $2062.66$ &  $1.4\pm0.5$ & $2.15\pm0.18$ & $-1.22\pm0.41$ & $0.013\pm0.005$ \\
CrII & $2066.16$ &  $5.4\pm2.5$ & $1.36\pm0.16$ & $-0.71\pm0.48$ & $0.010\pm0.005$ \\
FeII & $2249.88$ &  $325.3\pm44.5$ & $1.55\pm0.06$ & $-0.81\pm0.15$ & $0.027\pm0.004$ \\
FeII & $2260.78$ &  $313.6\pm33.5$ & $1.59\pm0.04$ & $-0.83\pm0.12$ & $0.035\pm0.004$ \\
FeII & $2344.21$ & * & $1.61\pm0.01$  & $-0.56\pm0.02$ & $0.353\pm0.006$ \\
FeII & $2374.46$ & * & $1.71\pm0.01$  & $-0.66\pm0.03$ & $0.179\pm0.005$ \\
FeII & $2382.77$ & * & $1.44\pm0.00$  & $-0.43\pm0.01$ & $0.533\pm0.006$ \\
MnII & $2576.88$ &  $2.4\pm0.2$ & $1.69\pm0.04$ & $-1.19\pm0.09$ & $0.052\pm0.004$ \\
FeII & $2586.65$ & * & $1.65\pm0.01$  & $-0.60\pm0.02$ & $0.317\pm0.005$ \\
MnII & $2594.50$ &  $1.7\pm0.2$ & $1.52\pm0.06$ & $-0.78\pm0.14$ & $0.028\pm0.004$ \\
FeII & $2600.17$ & * & $1.48\pm0.00$  & $-0.45\pm0.01$ & $0.534\pm0.006$ \\
MnII & $2606.46$ &  $2.0\pm0.3$ & $1.74\pm0.08$ & $-1.14\pm0.17$ & $0.024\pm0.004$ \\
MgII & $2796.35$ & * & $1.00\pm0.00$  & $0.00\pm0.00$ & $1.000\pm0.001$ \\
MgII & $2803.53$ & * & $1.11\pm0.00$  & $-0.01\pm0.01$ & $0.797\pm0.004$ \\
MgI & $2852.96$ &  $1.6\pm0.0$ & $1.45\pm0.01$ & $-0.67\pm0.03$ & $0.215\pm0.005$ \\
TiII & $3242.92$ &  $0.3\pm0.1$ & $1.55\pm0.15$ & $-0.15\pm0.38$ & $0.007\pm0.002$ \\
TiII & $3384.73$ &  $0.4\pm0.1$ & $1.51\pm0.09$ & $-0.31\pm0.26$ & $0.015\pm0.003$ \\
CaII & $3934.77$ &  $1.2\pm0.1$ & $1.53\pm0.03$ & $-0.77\pm0.12$ & $0.099\pm0.008$ \\
CaII & $3969.59$ &  $1.8\pm0.2$ & $1.57\pm0.06$ & $-1.32\pm0.20$ & $0.076\pm0.010$ \\
\hline
$\rm E_{g-i}$ & & $0.017\pm 0.003$ [mag] & $1.6\pm0.1$  & $-1.2\pm0.2$ & \\
$\rm E_{B-V}$ & & $0.011\pm 0.002$ [mag] & $1.6\pm0.1$  & $-1.2\pm0.2$ &\\ 

\hline
\end{tabular}

\begin{tablenotes}\footnotesize
\item $^{a}$ N column density values are estimated from the C parameter values in column 6 with Eq. 1.
\item $^{b}$ Saturated lines are marked with *

\end{tablenotes}
\end{threeparttable}

\label{table:all_data}
\end{table*}


%
\begin{figure*}
\center
\includegraphics[scale=0.35]{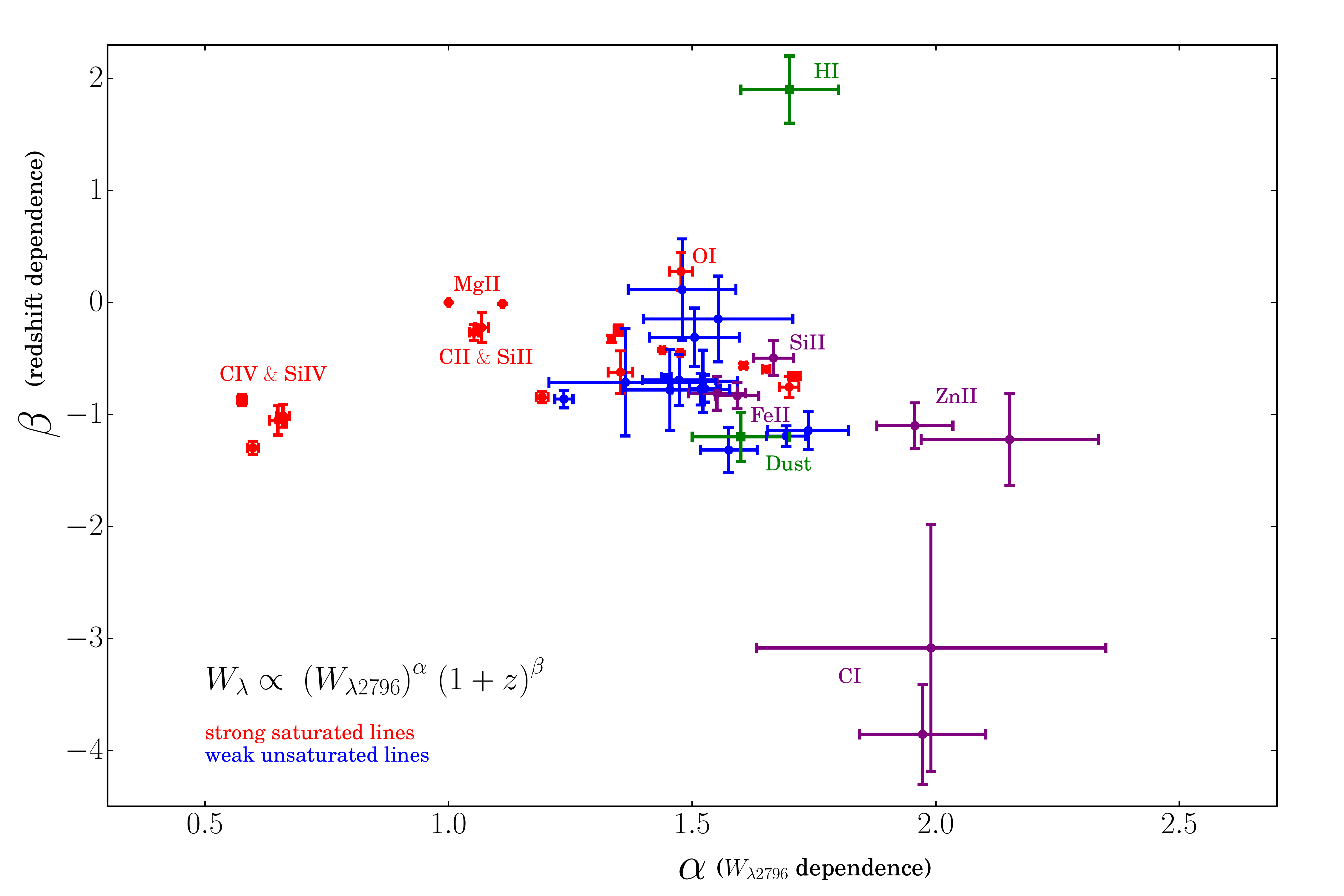}
\caption{Dependence of the equivalent width of each absorption line
  upon the strength of MgII absorption line (MgII$\lambda2796$),
  $\alpha$, and redshift, $\beta$, as parametrized in equation
  (3). Blue symbols indicate weak unsaturated lines, red 
  symbols are for saturated lines.  Purple symbols highlight Si II,
FeII, ZnII and C I which are discussed in the text.
Green symbols are used for
  the hydrogen column density and for extinction $E(B-V)$.}
\label{}
\end{figure*}
%

\section{Results}

\subsection{Evolution of the HI column density in MgII absorbers}
\label{HI_MgII_relation}
Let us first study neutral hydrogen in MgII absorbers.  
We take additionally Mg II absorber samples with neutral hydrogen column
densities measured with the Voigt fitting to study their HI content:
(i) The $z\sim 1$ sample from \citet{Rao2006}, which consists of 197
MgII absorbers at $z<1.7$ with $N_{\rm HI}\geq 10^{18}{\rm cm^{-2}}$;
(ii) high redshift ($z\sim 3$) sample from \citet{Matejek2013},
containing 33 systems with median redshift about 3.4. These are
supplemented by (iii) the recent low redshift ($z\sim 0.3$) sample of
16 Mg II absorbers in \citet{Rao2017}.

Figure 1 shows the samples in the order of blue symbols $(\langle z
\rangle\simeq0.6)$, green
$(\langle z \rangle\simeq1.2)$, and red $(\langle z
\rangle\simeq3.4)$, taken from the references cited above. We also show the median values of the samples with bootstrapping errors shown with square data points.  We observe
that the column densities of neutral hydrogen increase with the
equivalent widths of MgII absorption lines (MgII$\lambda2796$), and at
a fixed MgII absorption strength monotonically with redshifts, as
noted in \citet{Menard2009} and \citet{Matejek2013}.

These trends are summarised as
\begin{equation}
N_{\rm HI} = A \bigg(\frac{W_{\rm \lambda 2796}}{\rm 1 \AA}\bigg)^{\alpha} \bigg(1+z\bigg)^{\beta},
\end{equation}
where $W_{\rm \lambda 2796}$ refer to MgII$\lambda2796$.  From the fits to
three separate samples for different redshift bins we find that slopes for the
$W_{\lambda 2796}$ dependence are mutually consistent within errors.  We,
therefore, adopt a common parameter $\alpha$ for the slope for our full
sample analysis. To make use of all the information, we obtain the final best-fit parameter values through a fitting to all the individual data points ($\log \, N_{\rm HI}$, $W_{\lambda 2796}$, $z$)\footnote{We have obtained the best-fit parameter values with three methods,(1) fitting to all the individual data points, (2) fitting to median values with bins, and (3) fitting to mean values with bins, and confirmed that the three methods yield consistent best-fit parameters.}.

The neutral hydrogen column densities increase with the MgII
equivalent width with a power index $\alpha= 1.69\pm 0.13$.
For the other parameters we find $\beta = 1.88 \pm 0.29$, and $A=10^{18.96\pm0.10}{\rm
  cm^{-2}}$, which are also given in Table 1.  The errors are
estimated by bootstrapping the samples for 200 times.  The fits are
presented in Figure 1 for the three redshift bins, with median redshifts
0.6, 1.2, and 3.4.

\subsection{Metals in MgII absorbers}

Having measured the rest-frame equivalent width of each absorption line for
the composite spectrum for bins of subsample where 
the MgII equivalent width and the redshift specified, we fit 
to our data with
\begin{equation}
  W_{\lambda} = C \, \bigg(\frac{W_{\lambda 2796}}{1 \rm \AA}\bigg)^{\alpha} \bigg(1+z\bigg)^{\beta},
  \label{eq:Wfit}
\end{equation}
where $\alpha$, $\beta$ and $C$, are given in Table 1, and also
displayed in Figure 2, where the plots are basically divided into two classes,
saturated (red) and unsaturated lines (blue).  The green symbols
stand for the parameters for HI obtained above and that for dust
reddening parametrised in a similar way as to $\alpha$ and $\beta$,
i.e., $E(B-V)$ in place of $W_\lambda$ written in the form as equation
(\ref{eq:Wfit}) \citep{Menard2008,Menard2012}\footnote{The fit here
  refers to the rest-frame colour, as in \citet{Menard2008}, while a
  similar fit of \citet{Menard2012} uses the colour in the
  observed frame. The power indices of the two differ by 1.2,
  reflecting the $\lambda^{-1.2}$ dependence of the extinction curve.}. We show the measured rest equivalent widths and the best-fit functions in Appendix.

The following trends are seen in Figure 2:
\begin{itemize}

\item All metal absorption lines (with an exception of OI) show
  negative $\beta$, meaning an absorption stronger at lower redshift
  when measured at a fixed MgII equivalent width. This indicates the
  equivalent widths, or column densities in case of unsaturated lines,
  decrease approximately, in median, by a factor of 1.6 from redshift
  0.5 to 2. 

  This contrasts to neutral hydrogen, a larger equivalent width at
  higher redshift ($\beta\simeq1.9$). 

\item Most of the low-ionized, unsaturated absorption lines show
  $\alpha=1.4-1.7$ with a median $\sim 1.6$, which is consistent with
  $\alpha$ for $N_{\rm HI}$, Namely the abundances of metals in line
  of sight is proportional to neutral hydrogen column density, or in
  other words, metallicity of MgII clouds does not vary much with $W_{\lambda 2796}$.

  We remark that $\alpha$ of ZnII is somewhat larger, $\alpha\sim 2$,
  which is about 3 $\sigma$ away from the remainders.
  
  \item CI absorption shows both $\alpha$ and $\beta$ largely
    different from other lines. The $\beta\sim-3.9$ value indicates that the CI equivalent width decreases by more than a factor 5 from redshift 1
    to 2.5.
    We will pay a special attention to CI below.

\item Low-ionized saturated absorption lines, MgII, CII, SiII are
  nonevolving: $\beta\approx 0$. The $\alpha$ values being close to 1
  means a similarity to our reference MgII absorption line.  This
  is expected for the saturated rest equivalent width which
  is controlled by the velocity dispersion of the system.

\item Highly-ionized absorption lines, CIV and SiIV, are saturated.
  They show $\alpha \sim 0.6$ and $\beta\sim -1$ that differ from
  low-ionised saturated lines.

\item We added $E(B-V)$ for a reference, taken from
  \citet{Menard2012}.  $\alpha\simeq 1.6$ means $E(B-V)\propto N_{\rm
    HI}$, and $\beta\sim-1.2$ is consistent with the decrease of
  metallicity to higher $z$.
  
\end{itemize}

\begin{figure*}
\center
\includegraphics[scale=0.45]{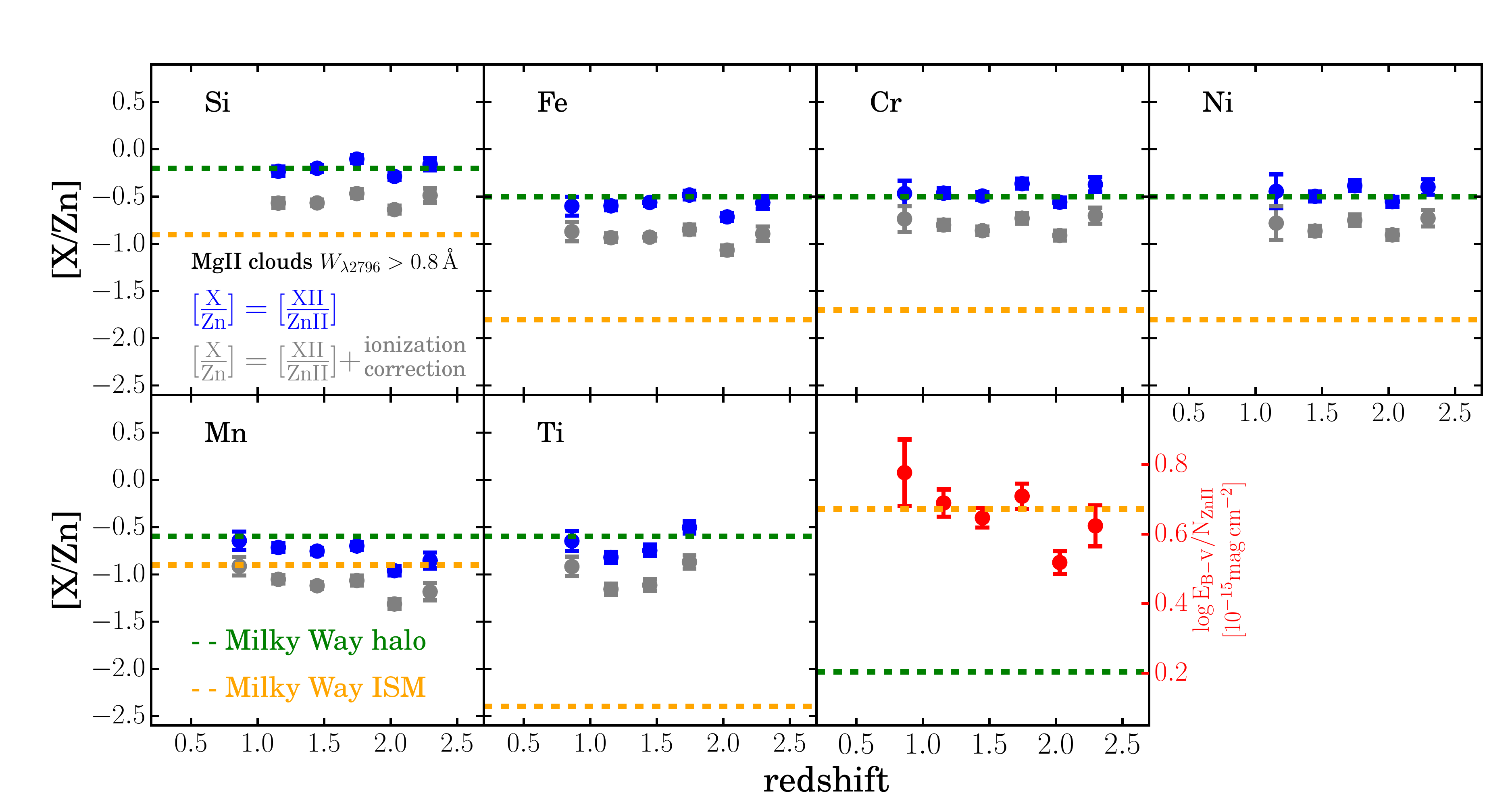}
\caption{Abundance of each element relative to Zn as a function
  of redshift. The last panel shows the dust abundance as estimated from
  $E(B-V)$ divided by the column density of ZnII. Horizontal lines
  indicate the corresponding value for Milky Way halo (green) and
  for interstellar matter (orange) \citep{York2006}. The zero point is the solar
  value.  Blue points refer to the relative abundance with no ionization correction, [X/Zn]=[XII/ZnII] and grey points refer to the relative abundance after the
ionisation corrections taken into account.}
\label{}
\end{figure*}

\begin{table*}[ht] 

\caption{Relative abundance patterns of MgII clouds, Milky Way halo, SMC, and Milky Way ISM}

\centering
\begin{threeparttable}
  \begin{tabular}{| c | c | c | c | c | c | c | c | c | c |}
\hline\hline
\hspace{0.4cm}

$\rm {[}X/Zn{]}={[}XII/ZnII{]}$                     & C      & Si   & Mg     & Fe   & Cr   & Ni   & Mn   & Ti   & Reference           \\
\hline\hline
MgII clouds                                & -      & -0.2 & -      & -0.5 & -0.5 & -0.5 & -0.7 & -0.6 & This paper \\
&&&&&&&&&\\
Milky Way halo                                & -0.2   & -0.2 & -0.2   & -0.5 & -0.5 & -0.5 & -0.6 & -0.6 & York et al. (2006)  \\
SMC (sk 108)                                  & -      & 0.1  & -      & -0.5 & -0.5 & -0.8 & -0.6 & -    & Welty et al. (2001) \\
&&&&&&&&&\\
Milky Way cool ISM                            & 0.1    & -0.9 & -0.8   & -1.8 & -1.7 & -1.8 & -0.9 & -2.4 & York et al. (2006)  \\
\hline 
\hline

$\rm {[}X/Zn{]}={[}XII/ZnII{]}+ionization \ correction^{a}$ &        &      &        &      &      &      &      &      &                     \\

MgII clouds                                & -0.6$^{b}$ & -0.6 & -0.6$^{b}$ & -0.9 & -0.9 & -0.9 & -1.1 & -1.0 &  This paper  \\      
\hline
\end{tabular}
\begin{tablenotes}\footnotesize
\item $^{a}$ The ionization correction for Zinc, $\log \, N_{\rm Zn}\approx \log \,N_{\rm ZnII}+0.4$, is estimated with $\rm log \, n_{H}=-0.5$.
\item $^{b}$ Assuming Milky Way halo values with ionization correction
\end{tablenotes}
\end{threeparttable}
\label{table:all_data}
\end{table*}

In Figure 3 we show the evolution of the abundance pattern for several
representative heavy elements in single ionized state in MgII absorbers
with $W_{\lambda 2796}>0.8 \rm \, \AA$, as a function of redshift, taking the
column density of ZnII as the reference. At this stage we do not apply
ionisation corrections, which turn out to be appreciable for Zn II, to
facilitate their comparison with elemental abundance in the
literature, in which ionisation corrections are usually not made.  The zero
point of the ordinate is the solar. Horizontal dashed lines show the
abundance taken from \citet{York2006} representing the value in
Milky Way's halo, which is similar to the SMC abundance \citep{Welty2001},
and that of cold interstellar matter. The figure
shows that the abundance in MgII clouds is similar to that in Milky
Way halo, in agreement with what was argued in York et al.  The
abundance is not in agreement with that of the interstellar medium of the Milky Way.
It deviates from the solar, but they become close if they are
shifted by 0.5 dex (0.2 dex for Si II) upwards. The relative abundance is also listed in Table 2. The upper part shows  values without ionization correction and the lower part lists values for MgII clouds with ionization correction taking into account, which will be discussed in Section 3.3 below.

The 0.5 dex smaller abundance of the iron group elements, is likely to
be ascribed to the condensation into grains. Si is also smaller than
solar by 0.2 dex, as is the abundance in the halo, which is also
ascribed to depletion onto grains. The metal abundance
relative to Zn evolves weakly in the redshift range we study.

\begin{figure}
\center
\includegraphics[scale=0.48]{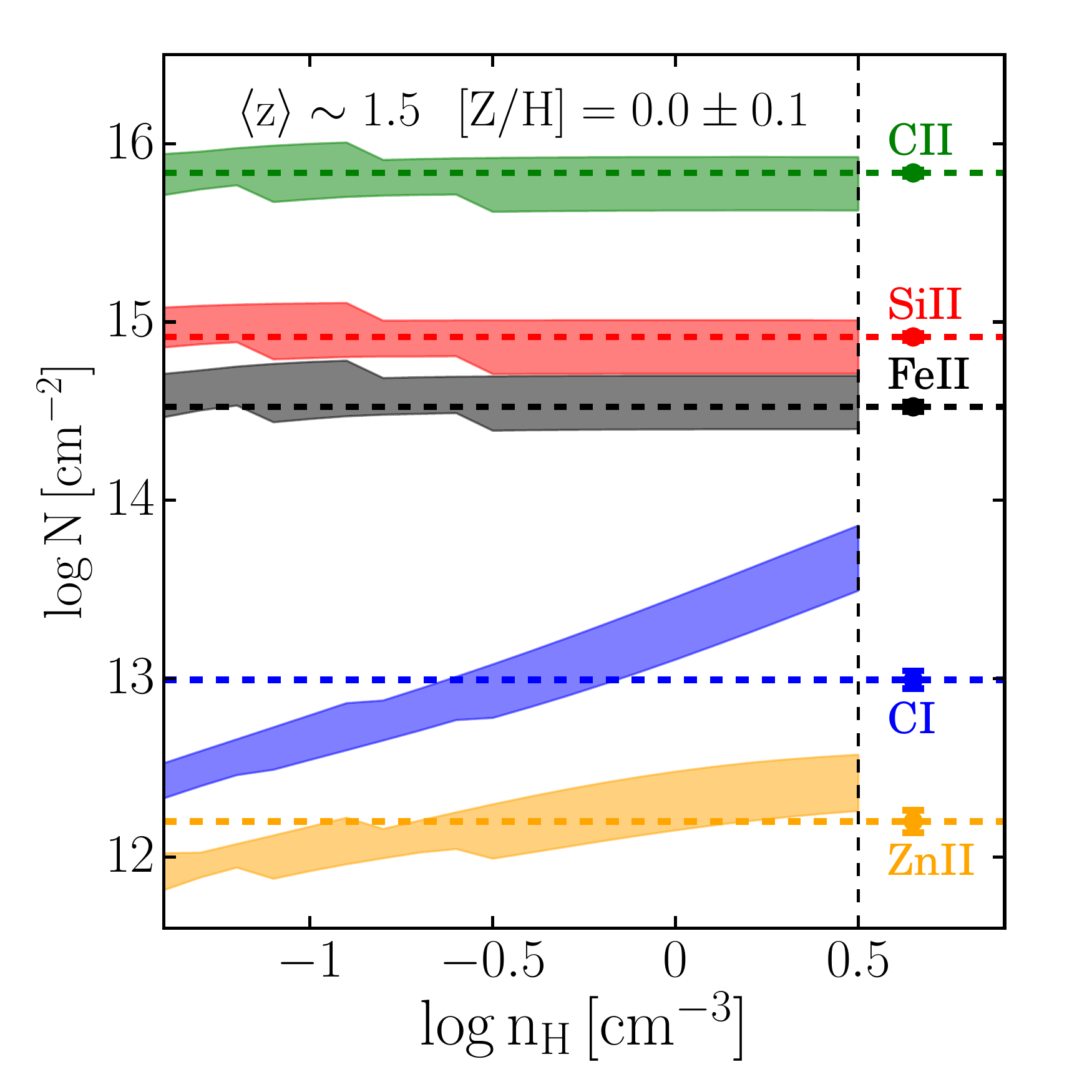}
\caption{Column density of various species: comparisons are made of
  the observed values (the symbols in the rightmost and horizontal
  lines extended therefrom) against the CLOUDY simulation plotted with
  varying hydrogen volume densities.}
\label{}
\end{figure}

The lower rightmost panel shows the dust abundance divided by the Zn
II abundance, where $E_{B-V}\simeq0.01 (W_{2796}/{\rm 1
  \AA})^{1.6}(1+z)^{-1.2}$ is used for the proxy for dust
\citep{Menard2008, Menard2012}.  This dust to metal in gas ratio is
close to the Milky Way's value \citep{Wild2006}, but is significantly
larger than the values in SMC or in Milky Way halo by a factor of~3.

\subsection{Physical conditions and ionisation corrections}

We have derived the column density of metals in their ionised
state. For Fe II, Si II etc., these are predominant ionisation states
and they represent practically the full column density of those heavy
elements. For other states, however, we must know the physical state
to infer how much fractions are in specific ionisation stages. When
compared with the observation, this in turn tells us about the physical
conditions of MgII absorbers.  The other uncertainty arises from
condensation into grains, as seen for iron and silicon.

We employ the CLOUDY
code\footnote{\url{http://trac.nublado.org/wiki}}\citep{Ferland2013}
to infer the ionisation correction factor for the given physical
condition.  We take the hydrogen column density $N_{\rm H}$, the
volume density of hydrogen $n_{\rm H}$, and metallicity of gas as
parameters.  
We set the
relative element abundance pattern to the Milky Way halo value listed in Table 2. 
We find that the ionisation correction for ZnII is appreciable, while it is not for FeII, MgII etc.

We assume that the cloud is photoionised by the background
radiation field, using the 2005 version of Haardt \& Madau
cosmic background radiation field
at each redshift, as in the default setting in CLOUDY\footnote{This is an unpublished
update
of their 2001 version \citep{Haardt2001}.
There is an
  alternative choice of the cosmic background radiation field
  \citet{Haardt2012}, which gives a photon spectrum somewhat tilted
  from their 2005 version. We have also carried out our analysis with the
  radiation field of \citet{Haardt2012}, which yielded metallicity by
  0.2$-$0.3 dex higher ($n_{\rm H}$ is lower by 0.3 dex). Because of this
  resulting supersolar metallicity, that looks unphysical, we do not take this
  2012 version as our setting.}.  
We constrain CLOUDY simulations with
the neutral hydrogen column density consistent with MgII
absorbers as observed, and explore $n_{\rm H}$ and metallicity that
reproduce the column densities of FeII, NiII, CI, Al III, CrII, ZnII,
and MgI as we observed. To constrain the physical conditions, we use measurements with high S/N derived from composite spectra with MgII absorbers with $W_{\lambda 2796}>0.8 \rm \, \AA$  in this section. We have also carried out the analysis as a function of $W_{\lambda 2796}$ and confirmed that the inferred physical conditions have a weak dependence with $W_{\lambda 2796}$.

Figure 4 shows an example of several metal column densities (SiII,
FeII, CI, CII, ZnII) at $z\simeq1.5$ from CLOUDY as a function of the gas
density $n_{\rm H}$, compared with the observed values.  Solar
metallicity [Z/H]=0 is assumed. The observed column densities are also
indicated by horizontal lines, together with the observation shown in
the right margin.  We see that the calculation becomes close to the
observation for almost all elements we consider when $-1<\log n_{\rm
  H}<0$.  We carry out this analysis at redshifts between $z=1$ and
2.5. We find that features of the curves of the metal column density
versus $n_{\rm H}$ change only at a quantitative level in this redshift
interval.  We note that CI is particularly sensitive to the hydrogen
volume density of the system, and also to redshift. Let us note that
CI is a minor component while CII is the predominant agent of carbon.

\begin{figure}
\center
\includegraphics[scale=0.48]{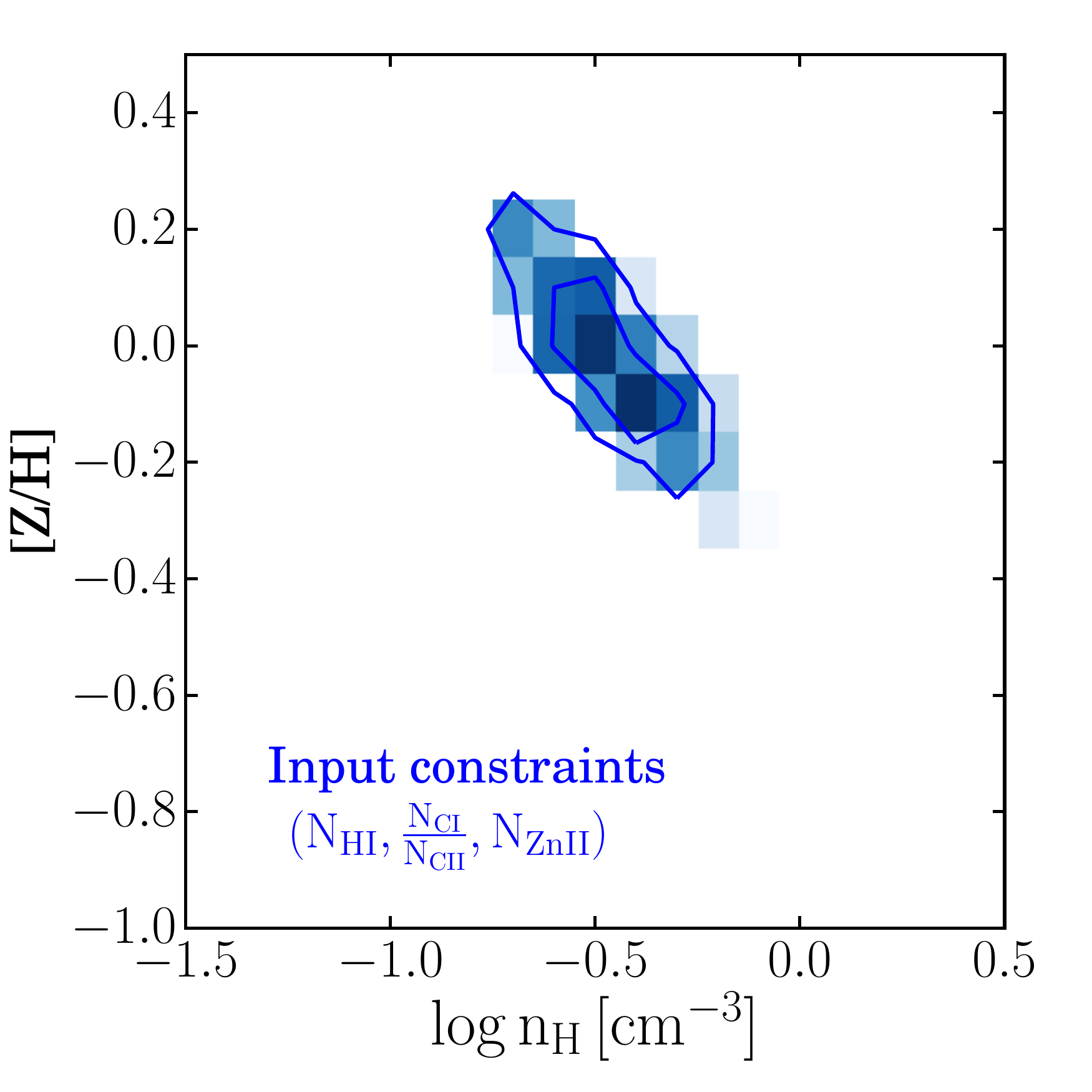}
\caption{Allowed region for metallicity [Z/H] and the gas density
$\log n_{\rm H}$ at redshift $z\simeq1.5$ constrained by the observed $N_{\rm HI}$, $\frac{N_{\rm CI}}{N_{\rm CII}}$, and $N_{\rm ZnII}$.}
\label{}
\end{figure}

We attempt to constrain simultaneously the volume density and the
metallicity with the observed ZnII column density and the ratio of
$N_{\rm CI}$ to $N_{\rm CII}$, which is estimated from $N_{\rm SiII}$,
as shown in Figure 5 for one and two sigma contours at $z\simeq1.5$. The
hydrogen volume density is about $\log n_{\rm H}\sim-0.5$ and
metallicity consistent with the solar.

In Figure 6 we show the redshift evolution of CI from CLOUDY, which
shows a rapid decrease with redshift, where we assume log $n_{\rm H}=-0.5$
for all redshifts.  The rapid evolution of CI agrees very well with
that we derived from the observation.

We then estimate the major component, CII abundance $N_{\rm CII}$ from
$N_{\rm CI}$ using the ionisation correction factor.  The CII column
density thus calculated is nearly constant in our redshift range,
where the ionisation correction factor for CI varies rapidly between
100 and 1000.  We compare it with $N_{\rm CII}$ estimated from $N_{\rm
  SiII}$ with the Milky Way halo abundance ratio of carbon to silicon.  The
agreement of the two estimates is impressive, verifying the validity
of CLOUDY results with the 2005 version of the Haardt \& Madau
ionisation field at each redshift.

We now lift the assumption of $\log n_{\rm H}=-0.5$ for all redshifts, and obtain the
best fit at each redshift bin using $N_{\rm HI}$, $N_{\rm CI}/N_{\rm CII}$ and $N_{\rm ZnII}$ as constraints.  Figure 7 shows log $n_{\rm
  H}=-0.5{+0.2 \atop -0.1}$ at all redshifts that concern us.  This indicates that a
strong redshift evolution of CI column density, $N_{\rm
  CI}\propto(1+z)^{-3.9}$, is induced most importantly by a decrease
of photoionising radiation field towards lower redshifts. This fast
evolution of CI column density accounts for a rapid evolution of the
CI cloud incidence, a rapid increase from z=2.5 to 1.5 observed in
\citet{Ledoux2015}.

We also study the prediction of Al III, ZnII and Mg I column
densities.  We find that the observed Al III column density is
consistent with the prediction for $-1\leq\log n_{\rm H}\leq -0.5$.  The
CLOUDY result for Zn II is consistent with the observation in so far as
$-1\leq \log n_{\rm H}\leq 0.5$. So, our choice, $\log n_{\rm
H}\approx -0.5$, from C I is a compromise, consistent with the
observation for Al III and Zn II.  On the other hand, the CLOUDY
calculation overestimates the observed MgI column densities by about
0.25 dex for $\log n_{\rm H}$ that concerns us.  There seems to be no
consistent value of $\log n_{\rm H}$.  This discrepancy suggests that
the input, including the background radiation field, may not fully
capture the relevant physics for Mg I, as also noted by \citet{Prochaska2017}.  The major components, Si II, Fe II, Cr II and Ni II, vary
little against $n_{\rm H}$ in so far as log $n_{\rm H}>-1$ and agree with the
observed values for a wide range of log $n_{\rm H}$ with the Milky Way halo abundance
composition as input.

\begin{figure}
\center
\includegraphics[scale=0.45]{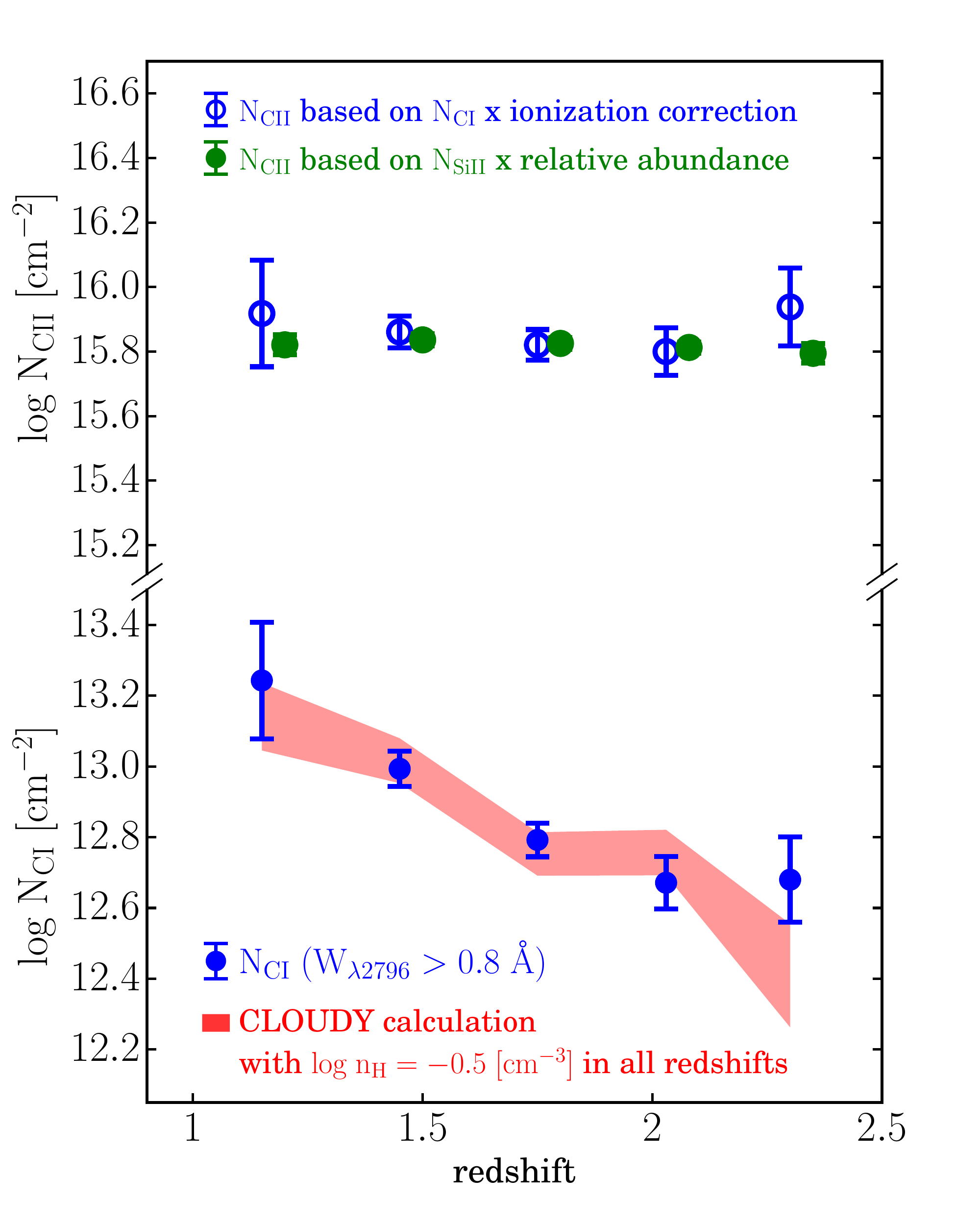}
\caption{Redshift evolution of the column density of neutral carbon in
  MgII absorbers: observed (blue symbol) versus CLOUDY simulation (red
  shades), shown in the lower part of the figure. In the upper part
CII column density calculated from CI with ionisation correction 
factor is shown together with CII from SiII using the Milky way halo abundance
composition.}
\label{}
\end{figure}

\begin{figure}
\center
\includegraphics[scale=0.5]{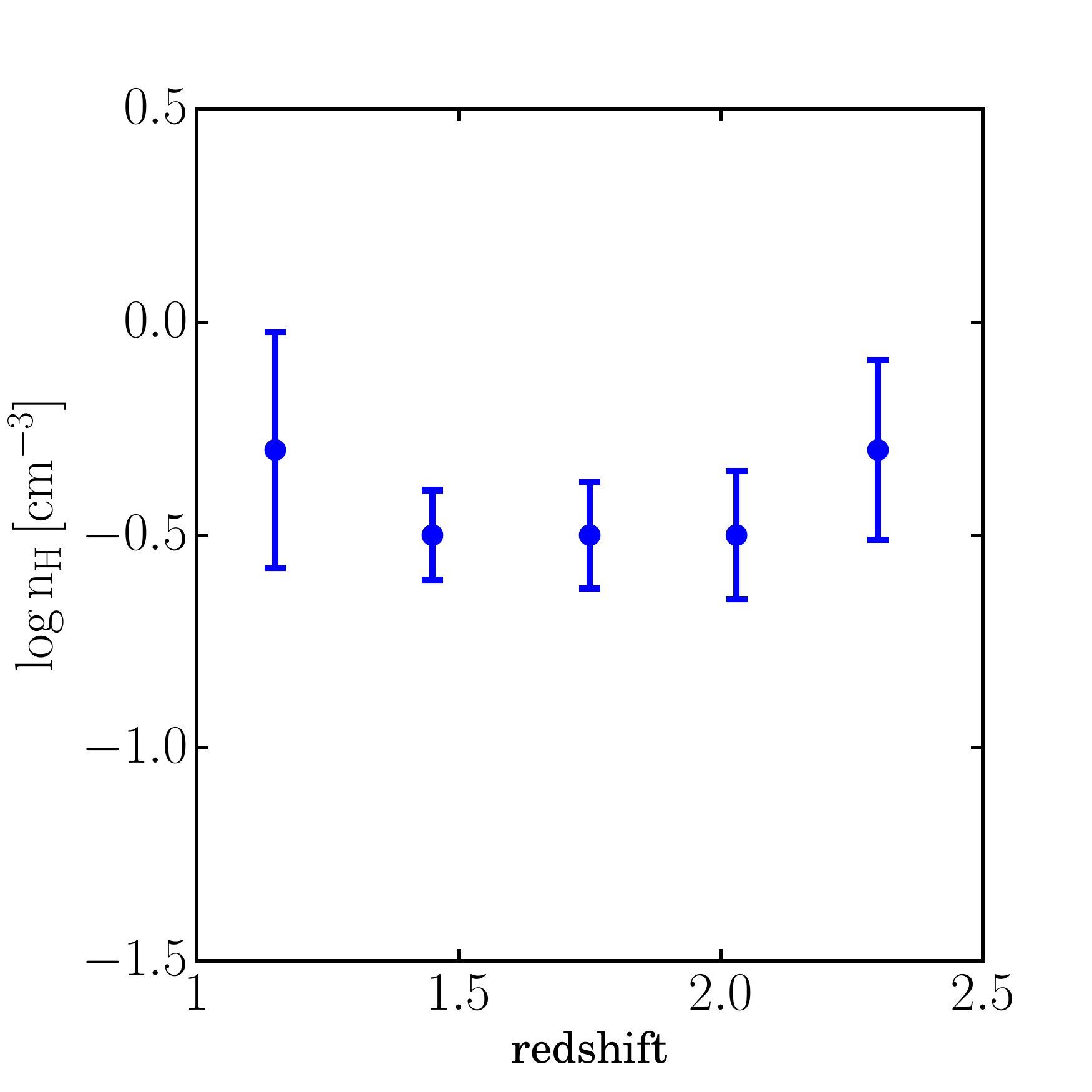}
\caption{Optimum values of the hydrogen volume density as a function of
  redshift.}
\label{}
\end{figure}

\begin{figure}
\center
\includegraphics[scale=0.45]{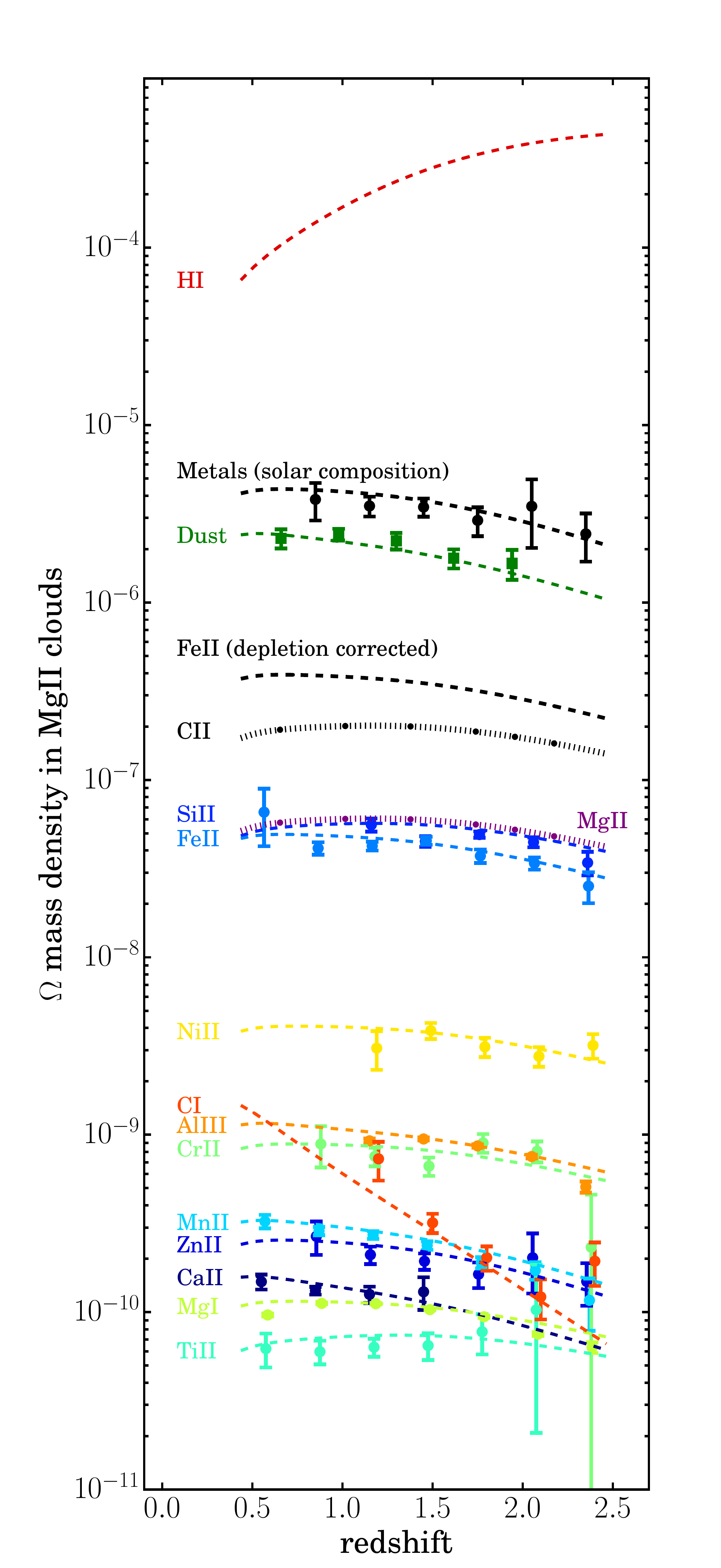}
\caption{Cosmic mass density of heavy elements and HI borne by MgII
  clouds. ${\rm Fe^{corrected}}$
means the curve with the depletion correction of 0.9 dex applied.  The
curve with 'metal' stands for the mass density of total heavy elements
obtained from Zn abundance after correction for the ionisation 
multiplied by solar
abundance of other heavy elements with respect to zinc. Other element
stands for spectroscopic estimates. Dust is taken from \citet{Menard2012}. Color dashed lines show the mass densities obtained by the best-fit parameters listed in Table 1.}
\label{}
\end{figure}

\begin{figure*}
\center
\includegraphics[scale=0.45]{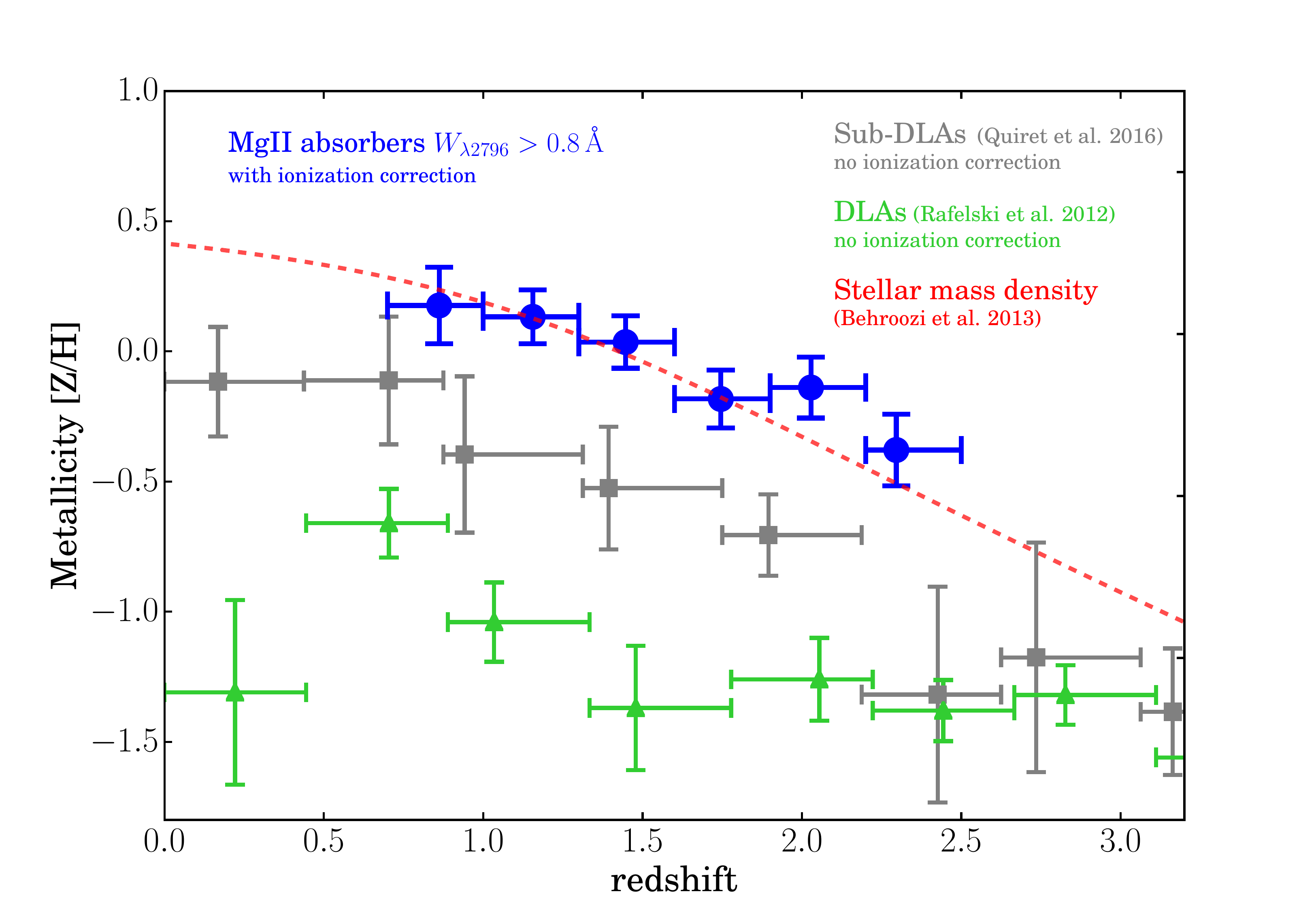}
\caption{Metallicity [Z/H] evolution. Metallicity of MgII clouds is
  shown with blue solid symbol. Metallicity of DLA is shown by green symbols
\citep{Rafelski2012}. 
Sub-DLA \citep{Quiret2016} is
represented with grey squares.  Dotted (red) curve shows the cumulative
amount of stars born drawn with an arbitrary normalisation.}
\label{ZHevolution}
\end{figure*}

We estimate the heavy element abundance relative to Zn taking into
account the ionisation correction derived from CLOUDY with $\log n_{\rm H}=-0.5$, which affects notably Zn estimated
from ZnII: $\rm \log N_{Zn}\approx \log N_{ZnII}+0.4$. We showed the resulting Z/Zn
in Figure 3 above. After ionisation corrections for ZnII, the depletion of the
iron group elements becomes (e.g., $\rm [Fe/Zn]\approx[FeII/ZnII]-0.4$) $\sim0.9$ dex, rather than 0.5 dex, and that for silicon is $\sim0.6$ dex, rather than 0.2 dex: see grey symbols in the
figure and the lower part of Table 2.

We find from CLOUDY that $n_{\rm HI}/n_{\rm H}=0.8-0.9$ for $\log n_{\rm H}=-0.5$, hydrogen predominantly being neutral.  The inferred temperature of the cloud
is about 2500 K with about 300 K uncertainty .  The neutral fraction stays at greater than 0.5 unless $n_{\rm H}$
becomes smaller than $-1.5$, for which temperature goes up to greater than 5000 K.
These results change weakly with redshift.

The derived volume density of HI has a significant implication
concerning size of the MgII cloud.  It implies that cloud sizes are of the order of $ r_{\rm cloud}\sim N_{\rm
  HI}/n_{\rm HI}\sim 0.03$ kpc, much smaller than the size of
galaxies. This is compatible with the spatial size
   inferred for one specific cloud, showing Si II and CII absorption
   features, from a gravitationally lensed quasar \citep{Rauch1999}.
This is also compatible with the size of the cloud derived
earlier \citep[e.g.][]{Rigby2002, ProchaskaHennawi2009, Crighton2015}.  For
this size we infer the typical baryonic mass of MgII clouds of the
order of $M_{\rm cloud} \sim \rm 10^{3} \ M_{\odot}$.  Considering the
fact that the covering factor of Mg II clouds around galaxies at redshift $\sim$0.5 are
typically 0.5 \citep[e.g.,][]{Chen2010,Menard2012,Nielsen2013,Lan2014}, at a distance typically $20-50$ kpc
from the galaxy centre.  This size means that Mg II clouds that
surround galaxies should be numerous, say, at least $10^5-10^6$. So,
they are like patchy clouds or foam-like objects that surround
galaxies.  This is also consistent with multiple components of Mg II
clouds in the velocity space seen in many sight lines to quasars, as
observed in \citet{Churchill2003}.

\subsection{Cosmic mass density of metals}

The cosmic mass density of HI in MgII clouds is estimated from
\begin{eqnarray}
\rho_{\rm HI}^{\rm MgII}(z) = \frac{m_{\rm HI}}{dX/dz} \int_{W_{\rm min}}^{\infty}dW_{\lambda 2796}\,\frac{dN}{dW_{\lambda 2796}\,dz} \times
\nonumber\\
N_{\rm HI}(W_{\lambda 2796},z)
\label{eq:density}
\end{eqnarray}
where
${dN}/{dW_{\lambda 2796}\,dz}$ is the incidence rate of MgII absorbers
(taken from \citet{Zhu2013}),
and $N_{\rm HI}(W_{\lambda 2796},z)$ is the HI column density we derived
in equation (2): $X$ is the absorption distance.  We take $W_{\rm
  min}=0.4$ \AA~ as a default.  We study the convergence of the
integral towards the weak line limit. We find 30\% decrease of
$\Omega_{\rm HI}$ if $W_{\rm min}$ is increased to $0.8$ \AA. Our
extrapolation to $W_{\rm min}=0$ indicates the increase of the
integral from the 0.4 \AA~ cutoff to be at most a few percent in
$\rho_{\rm HI}$: the integral is fairly well convergent with our
default $W_{\rm min}$.  The HI mass density obtained in equation
(\ref{eq:density}) decreases towards zero redshift, as shown in Figure 8:
 $\Omega\simeq 4\times 10^{-4}$ at $z=2$ decreases to $1\times
10^{-4}$ at $z=0.5$.  This estimate is consistent with
\citet{Menard2012}.

We estimate the mass density of various elements in respective
ionisation stages, or the density of species after the ionisation
correction, by replacing HI with the relevant element and state in
equation (\ref{eq:density}). In Figure 8 we show the cosmic mass
density residing in the MgII cloud for various species, CII, MgII,
SiII, FeII, NiII, CI, AlIII, CrII, MnII, ZnII, CaII, MgI and TiII.  
For most of the specific elements, those ionisation stages depicted in the figure
stand for the predominant state in the cloud.  We note that the curve
for SiII is degenerated with that of MgII, including the redshift
dependence. We also add a curve corrected for the depletion for Fe, total heavy element abundance
(Metals) in MgII clouds taking the solar composition, and dust abundance in MgII clouds from \citet{Menard2012}.

The cosmic density of Fe from our spectroscopic analysis after the 0.9
dex depletion correction is about $\Omega_{\rm Fe}^{\rm MgII}\simeq
4\times 10^{-7}$ at $z\approx 1.5$, as seen in Figure 8.  
The 0.9 dex Fe depletion observed in gas phase means that about $90\%$ of Fe ($\Omega_{\rm Fe}^{\rm Dust} \sim 3.6 \times 10^{-7}$) is locked in dust grains if the origin metal composition is assumed to be solar.
It would be interesting to compare this amount with the estimate from dust in MgII absorbers. 
In \citet{Menard2012} the amount of dust in MgII absorbers is
estimated to be $\Omega_{\rm dust}^{\rm MgII}=2.0\times 10^{-6}$
($z=1.5$) from dust reddening behind MgII absorbers using broad-band
quasar photometry (green data points). With typical iron fraction
Fe/dust$\approx 0.2$ \footnote{We take dust to consist of 70\%
  astronomical silicate and 30\% graphite \citep[e.g.][]{Draine2011}.} we find $\Omega_{\rm
  Fe}^{\rm Dust}\approx4\times 10^{-7}$ in agreement with our
spectroscopy-based estimate.

The increase of mass density of metals, as borne by MgII clouds
is moderate towards $z=0$. We note, however, that the HI mass density
is significantly decreasing towards lower redshift, by a
factor of 4 from $z=2$ to $z=0.5$.  This means a
loss of material in Mg II clouds, say for example by their falling on 
galaxies or destruction for some reasons. This means that
heavy elements in Mg II clouds 
are most likely lost along with HI gas. 
In other words, if we
correct for this loss factor, the evolution of metal abundance should
be traced by Z/H, rather than Z: the increase of the metal abundance
towards lower redshift should then be significant.

The same comment also applies to the dust abundance in \citet{Menard2012}, which shows a slow evolution with redshift. The cosmic evolution of dust should be obtained by dividing their values with the
mass density of HI, correcting for the gas mass loss in Mg II clouds towards
low redshift.

\subsection{Evolution of the global metal abundance}

We study the metallicity evolution traced by MgII absorbers.
Namely the measure of metallicity,
Z/H, stands for the abundance of heavy element that would be contained
in MgII clouds. When a correction is taken into account for the
redshift evolution of hydrogen mass density in MgII clouds, we are led
to the global abundance of metals.

The resulting [Z/H] (denoted by blue solid circles in Figure 9) 
evolves from redshift 2.5 to 0.5, increasing by a factor of 4.
This quantity can be interpreted as ordinary metallicity
[Z/H]. It reaches close to solar, in fact, approximately twice the
solar metallicity at zero redshift.

In this figure we compare [Z/H] of Mg II absorbers with other cool gas
absorbers. The similar pattern of the evolution seen with MgII
absorbers is seen in that of DLAs (green points) \citep{Rafelski2012}
and of sub-DLAs (grey points) \citep{Quiret2016}\footnote{In this figure all errors are by bootstrapping.
For DLA and subDLA data points we estimate the
median metallicity of their samples with bootstrapping error bars
of metallicity instead of the hydrogen weighted metal abundances.}. 
It is important to note that the ionisation correction was not usually
applied to the data of DLA and sub-DLA. The correction would raise
those curves approximately by 0.4 dex for sub-DLA \citep[see also][]{Som2015}, for which we expect the physical condition is similar to Mg II clouds. In our result of MgII the ionisation
correction, most importantly that for ZnII, is included that shifted
the curve upwards by about 0.4 dex.

Metallicity of Mg II clouds, with or without the ionisation correction, is
larger than that of DLA.  It is only about 0.2 dex larger than
metallicity of sub-DLA, but larger by $\approx$ 1 dex than that
of DLA.  This approximately agrees with \citet{Fukugita2015}, which
shows that metallicity is inversely proportional to $N_{\rm HI}$ in
DLA.

We show in this figure the cumulative amount of fuel used for the
star formation rate in galaxies (the star formation rate is taken from
\citet{Behroozi2013} integrated from a high to the relevant redshift
(red dashed line) with an arbitrary normalisation. It is interesting
to observe that [Z/H] of MgII clouds, i.e. what traces the total
amount of metals in intergalactic media, closely traces the cumulative
fuel consumed to that redshift.  Metals are produced in stellar
evolution: if a constant fraction of metals produced in stars is
transported to intergalactic space by the galactic wind, it will
contaminate circumgalactic space. Hence, it is natural to suppose that
metallicity in circumgalactic space, and so in Mg II clouds, is
proportional to cumulative star formation in galaxies
\citep{Menard2012}.

We note that the metal abundance shown in Figure 8 above means
   that contained in Mg II clouds. We must correct for the loss of gas mass 
   in the Mg II clouds towards low redshift to obtain the global
   metal density evolution. Namely we must divide each component of $\Omega_i$
   by $\Omega_{\rm HI}$.
\section{Summary}

Large data bases of quasar absorption lines have enabled a
spectroscopic study of the elemental abundance by stacking weak,
unsaturated lines, such as Fe II, Zn II, C I, Si II, Ni II and so on, of
many quasars as a function of line strengths of the Mg II lines but also of redshift.

We find that the abundance pattern of Mg II clouds resembles that of
Galactic halo or of SMC, but it differs significantly from solar or
that of Galactic interstellar matter, as has been inferred from the
extinction curve \citep{York2006,Fukugita2015}.  We find, however,
that the total abundance of heavy elements is larger than that of SMC
and is close to that of Milky Way.  We also confirm that iron group
elements are significantly (approximately 0.9 dex) depleted in cloud
spectra, as is known in interstellar gas or in
DLA \citep{vanSteenberg1988, Pettini1994}.  On the other hand,
depletion of zinc is, if any, not significant, since the zinc
abundance, or when it is multiplied with Z/Zn, is extrapolated to
solar at $z=0$. Therefore, we have taken zinc as our reference for metals,
in agreement with earlier reports, however after the ionisation
correction for Zn II is taken into account. The ratios, Fe/Zn, Si/Zn,
etc., which are consistently smaller than solar due to depletion,
evolve weakly with redshift.

We find a significant evolution of metallicity Z/H in MgII
clouds from our highest $z=2.5$ to the lowest $z=0.5$, an increase by
a factor of 4.  We find that this is mainly caused by a decrease of
total HI column density or the HI abundance of clouds towards lower
redshift.
We argue that the evolution of cosmic metal abundance reflects
in Z/H with the denominator taking account of the evolution of neutral hydrogen in Mg II clouds. 
Figure 9 shows that the evolution of Z/H closely traces the cumulative amount of hydrogen
fuel used for star formation in galaxies: evolution of metals
in Mg II clouds reflects star formation of galaxies.

Among the heavy elements we studied the species that shows a rapid
evolution is C I, which exhibits an increase by a factor of a dex
towards a low redshift ($z=0.5$) in its column density, caused mainly
by the decrease of the ionising radiation. This accounts for a rapid
increase of the CI cloud incidence towards low redshifts reported in
\citet{Ledoux2015}.

Net HI gas bound in the MgII clouds evolves from $\Omega_{\rm
HI}=4\times 10^{-4}$ at $z=2$ to $1\times 10^{-4}$ at $z=0.5$. We
also estimate the mass density of various elements. In
addition, we show that the iron abundance in dust inferred from the
depletion in gas phase $\Omega\approx 4\times 10^{-7}$ agrees with the
iron abundance inferred from dust in Mg II clouds using reddening of
quasars behind the cloud.

With the aid of the CLOUDY code we infer that the volume density of
the gas is roughly $\rm 0.3 \, cm^{-3}$, which does not vary with
redshift. This conclusion rests on the validity of CLOUDY
calculations, but we are convinced with its reliability from the fact
that it infallibly gives reasonable elemental abundance in so far as
we have tested. In particular, the carbon abundance estimated from
largely redshift-dependent CI becomes consistent with a constant after
the use of the ionisation correction factor of CLOUDY, and the
carbon abundance agrees with the predominant component CII,
inferred from the SiII abundance.  Moreover, with this density, the
evolution of the CI column density as given by CLOUDY shows nearly a
perfect match with that observed in our redshift range. The $n_{\rm H}$
dependence of ZnII, and Al III also match well between CLOUDY and the
observation with this gas density, and we find $n_{\rm H}\approx0.3 \, {\rm
  cm^{-3}}$ is a compromise to account for ZnII and Al III.

The abundances in single ionized metal elements, such as Si II, Fe II, Ni II, etc, are also in a good agreement between CLOUDY and our observation with the Milky Way halo/SMC abundance pattern.  These calculations also tell us that
hydrogen in the MgII clouds is predominantly ($\approx 80-90$\%) in
the HI state and the temperature is estimated to be roughly 2500 K.
Our elemental analysis overall seems to verify a validity of CLOUDY in
our problem. We stress that C I serves as a sensitive indicator for the
physical condition.

Our derived volume density of gas $\rm 0.3 \, cm^{-3}$ implies, together
with a typical column density of the clouds $3\times10^{19}{\rm
  cm}^{-2}$, that the size of clouds being 0.03 kpc, which is
compatible with earlier inference \citep{Rauch1999,Rigby2002,
  ProchaskaHennawi2009, Crighton2015}. If this is a typical size in
one dimension, numerous, say $10^6$, clouds are needed to explain the
observational indication that the covering factor of Mg II clouds
around galaxies amounts to 50\% of the sky.  This means that Mg II
clouds are like foam that surrounds the galaxies.  This picture would
explain what was found in spectroscopic observation showing
multi-components of the cloud in many line of sight
\citep{Churchill2003}. Typical baryonic mass of Mg II clouds is of the
order of $10^3 \, \rm M_\odot$.

\acknowledgements
We thank Brice M\'enard and Guangtun Zhu, who made their MgII catalogue
and absorption spectra available to us in a digital form.
MF thanks Hans B\"ohringer and Yasuo Tanaka for the hospitality at the
Max-Planck-Institut f\"ur Extraterrestrische Physik and also Eiichiro
Komatsu at Max-Planck-Institut f\"ur Astrophysik, in Garching. He also
wishes his thanks to Alexander von Humboldt Stiftung for the support
during his stay in Garching, and Monell Foundation in Princeton at
Institute for Advanced Study.  He received in Tokyo a Grant-in-Aid
(No. 154300000110) from the Ministry of Education.  Kavli IPMU is
supported by World Premier International Research Center Initiative of
the Ministry of Education, Japan.

\appendix

This appendix describes the details of our analysis discussed in
Section 3.1. To detect and measure weak unsaturated lines, we first
make median composite spectra by grouping the sample in absorber redshift and
$W_{\lambda 2796}$ bins. Figure 10 shows a composite spectrum obtained by
combining 8097 individual spectra from absorbers at $1.6<z<1.9$ and
with $W_{\lambda 2796}>0.8 \,\rm \AA$. We show two plots, black
the composite spectrum and blue the same plot with an enlarged scale, with
the scales shown on the left and right, respectively. We note that the
signal to noise ratio of the composite spectrum is about 500-1000 and
this high S/N enables us to detect weak metal lines as labeled in the
figure.

The rest equivalent widths of metal lines are measured from composite
spectra with Gaussian profile fittings, as our fits shown in Figure
11.  Isolated metal lines are measured with a single Gaussian profile
fit, while for wavelength regions with multiple lines we adopt
multiple Gaussian profiles to simultaneously fit all lines in the
vicinity.

Finally, in Figure 12, we show the measured rest equivalent widths of
all the metal absorption lines grouped with $W_{\lambda 2796}$ as a
function of redshift. The rest equivalent widths of the metal lines
increase with $W_{\lambda 2796}$ (from cyan to purple) and evolve
modestly with redshift (except for CI lines). We describe these
dependences with Equation 3 and show the best-fit functions with colour
lines in the figure. The corresponding best-fit parameters are listed
in Table 1 and shown in Figure 2 in the main text.


\begin{figure*}
\centering
\includegraphics[scale=0.24]{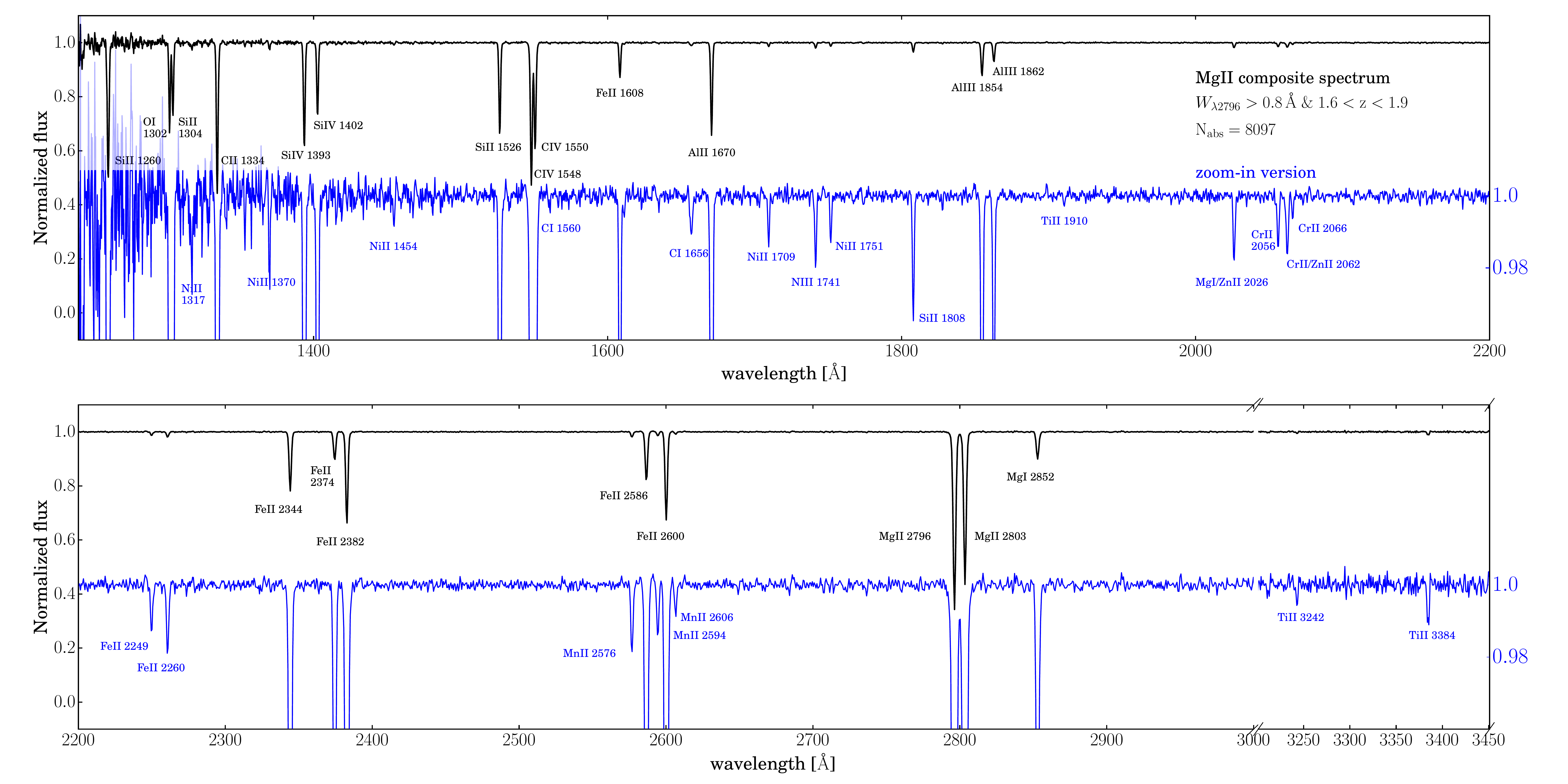}
\caption{Example of composite spectrum. It is obtained by combining 8097 individual spectra from MgII absorbers at redshift $1.6<z<1.9$ and $W_{\lambda 2796}>0.8 \,\rm \AA$. The black and blue spectra are identical and shown to emphasize absorption features in two scales. The y-axis scales for black and blue spectra are shown on the left and right, respectively. }
\label{}
\end{figure*}

\begin{figure*}
\center
\includegraphics[scale=0.6]{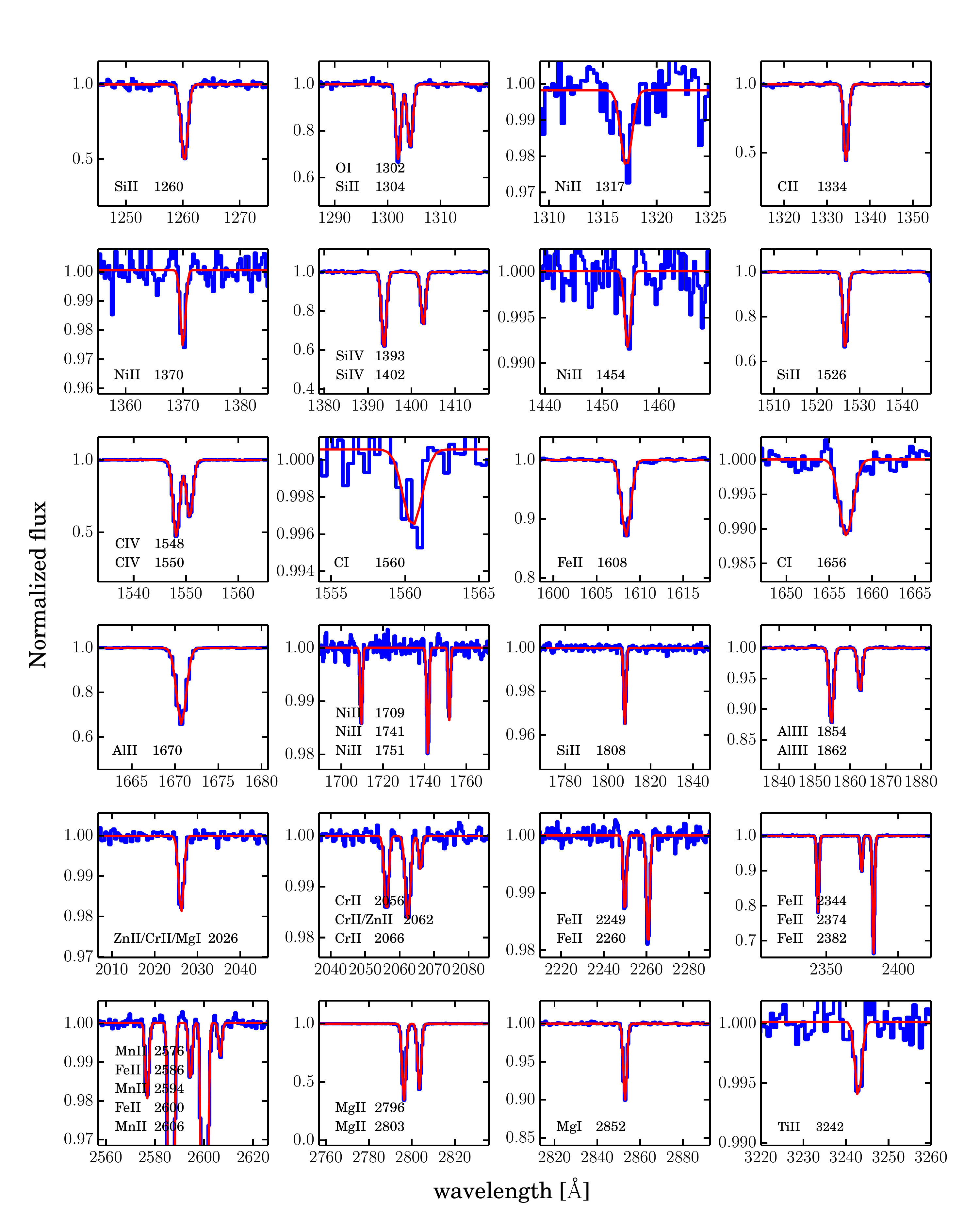}
\caption{Example of the gaussian fittings for metal absorption lines. The observed composite spectrum is in blue and the red curves are the best-fit Gaussian profiles.}
\label{}
\end{figure*}
\begin{figure*}
\center
\includegraphics[scale=0.35]{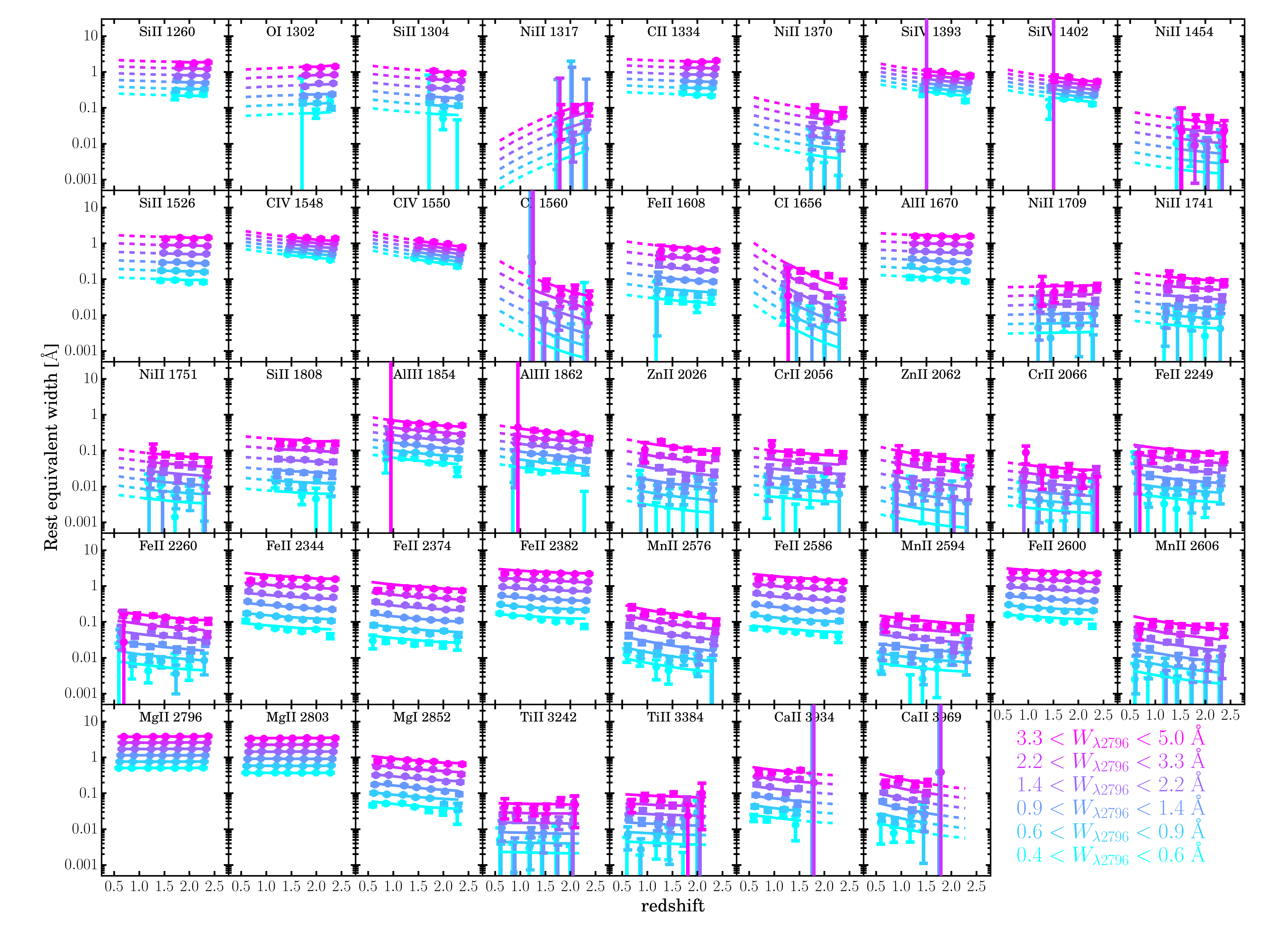}
\caption{Rest equivalent widths of 43 metal absorption lines as a function of redshift and $W_{\lambda 2796}$. The color indicates $W_{\lambda 2796}$ from weak (cyan) to strong (purple). Best-fit functions with Equation 3 are shown with color lines and the parameter values are listed in Table 1.}
\label{}
\end{figure*}

\end{document}